\documentclass[aps,10pt,pre,superscriptaddress,twocolumn,sort&compress,round,footinbib]{revtex4}

\usepackage{graphicx}
\usepackage{amsmath, amssymb}
\usepackage{gensymb}
\usepackage{dcolumn}
\usepackage{textgreek}
\usepackage{xcolor}
\usepackage{graphicx,bm,epsfig}
\usepackage{epstopdf}
\usepackage[english]{babel}
\usepackage{hyperref}
\hypersetup{
    colorlinks = true,
    citecolor = blue,
    linkcolor = blue,
    filecolor = magenta,
    urlcolor = blue,
    hypertexnames = true,
}
\definecolor{orange}{rgb}{0.8500, 0.3250, 0.0980}
\definecolor{amber}{rgb}{1.0, 0.75, 0.0}
\definecolor{arsenic}{rgb}{0.23, 0.27, 0.29}
\definecolor{battleshipgrey}{rgb}{0.52, 0.52, 0.51}
\definecolor{charcoal}{rgb}{0.21, 0.27, 0.31}
\definecolor{darkelectricblue}{rgb}{0.33, 0.41, 0.47}
\definecolor{firebrick}{rgb}{0.7, 0.13, 0.13}
\definecolor{azure}{rgb}{0.0, 0.5, 1.0}
\definecolor{purple}{rgb}{0.63, 0.36, 0.94}
\newcommand{\SImum}{\textrm{\textmu{}m}}
\begin{document}

\title[Transition from clogging to continuous flow in constricted particle suspensions]{Transition from clogging to continuous flow in constricted particle suspensions}

\author{Mathieu Souzy}
\email{m.p.j.souzy@utwente.nl}
\affiliation{Physics of Fluids, University of Twente, The Netherlands}

\author{Iker Zuriguel}
\email{iker@unav.es}
\affiliation{Departamento de F\'isica, Facultad de Ciencias, Universidad de Navarra, Spain.}

\author{Alvaro Marin}
\email{a.marin@utwente.nl}
\affiliation{Physics of Fluids, University of Twente, The Netherlands}

\date{\today}
\setcounter{page}{1}

%============================================================
%	Abstract
%============================================================

\begin{abstract}

When suspended particles are pushed by liquid flow through a constricted channel they might either pass the bottleneck without trouble or encounter a permanent clog that will stop them \emph{forever}. {\color{black} However, they may also} %But they might as well 
flow intermittently with great sensitivity to the neck-to-particle size ratio $D/d$. In this work, we experimentally explore the limits of the intermittent regime for a dense suspension through a single bottleneck as a function of this parameter. To this end, we make use of high time- and space-resolution experiments to obtain the distributions of arrest times {\color{black} ($T$)} between successive bursts, which display power-law tails ({\color{black} $\propto T^{-\alpha}$}) with characteristic exponents. These exponents compare well with the ones found for as disparate situations as the evacuation of pedestrians from a room, the entry of a flock of sheep into a shed or the discharge of particles from a silo. Nevertheless, the intrinsic properties  of our system {\color{black}(i.e. channel geometry, driving and interaction forces, particle size distribution)} seem to introduce a sharp transition from a clogged state {\color{black} ($\alpha\leq2$)} to a continuous flow, where clogs do not develop at all. This contrasts with the results obtained in other systems where intermittent flow, with power-law exponents above two, were obtained.

\end{abstract}
\maketitle

%===============================================================================
%	INTRO
%===============================================================================
%\section{Introduction}\label{Intro}

When people, animals, or particles are forced through a constriction, the flow may become intermittent due to the development of clogs that obstruct the constriction. Despite the diverse nature and scale of these systems -- including hungry sheep herds \cite{Zuriguel2014, Garcimartin2015sheep}, pedestrian crowds trying to escape a room in a life-and-death situation \cite{Helbing2005, Krausz2012}, discharge of dry granular silos \cite{Janda2009, Zuriguel2003, To2001, Mankoc2009}, suspended hydrated particles transported through pipelines \cite{Sloan2011}, electrons passing through nanoconstrictions on a liquid helium surface \cite{Rees2012}, %microfluidic and filtration circuits where a loaded liquid enters a device or permeates a membrane \cite{Bacchin2011},
sand hourglasses \cite{Wu1993}, or mice escaping a water pool \cite{Saloma2003} -- a distinctive phenomenology is always observed: particles or bodies flow in erratic bursts separated by short period of arrest. The analogy does not seem to be only qualitative and in all cases the number of escapees per burst follows an exponential distribution, and the probability distribution of time lapses separating the passage of consecutive bodies seems to exhibit a power-law tail with characteristic exponents that depend on diverse system parameters. These common features lead Zuriguel \emph{et al.} to propose a unified description \cite{Zuriguel2014}, providing a general framework for the study of intermittent clogging systems.

\begin{figure}[!h]
\begin{center}
	\includegraphics[width=0.5\textwidth]{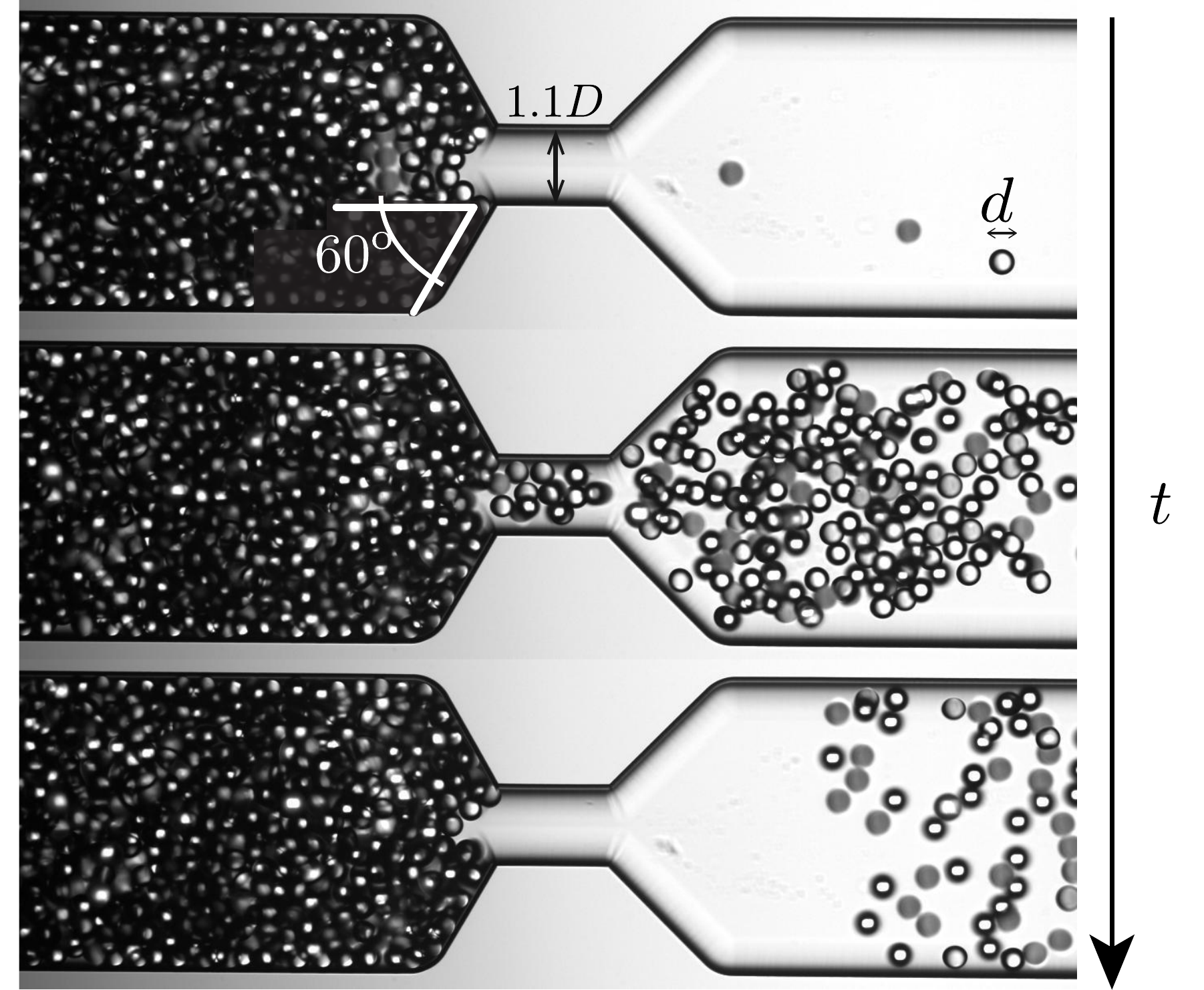}
	\caption{Successive snapshots of a burst in a suspension of particles with a diameter $d$ that intermittently flow through a constriction -- having a neck width $1.1D$ and height $D$ -- by a volume-driven flow ($D/d=3.03$). From top to bottom: clog previous to the burst, burst, and clog after the burst.}
	\label{fig:setup}
\end{center}
\end{figure}

\begin{figure*}
	\includegraphics[width=1\textwidth]{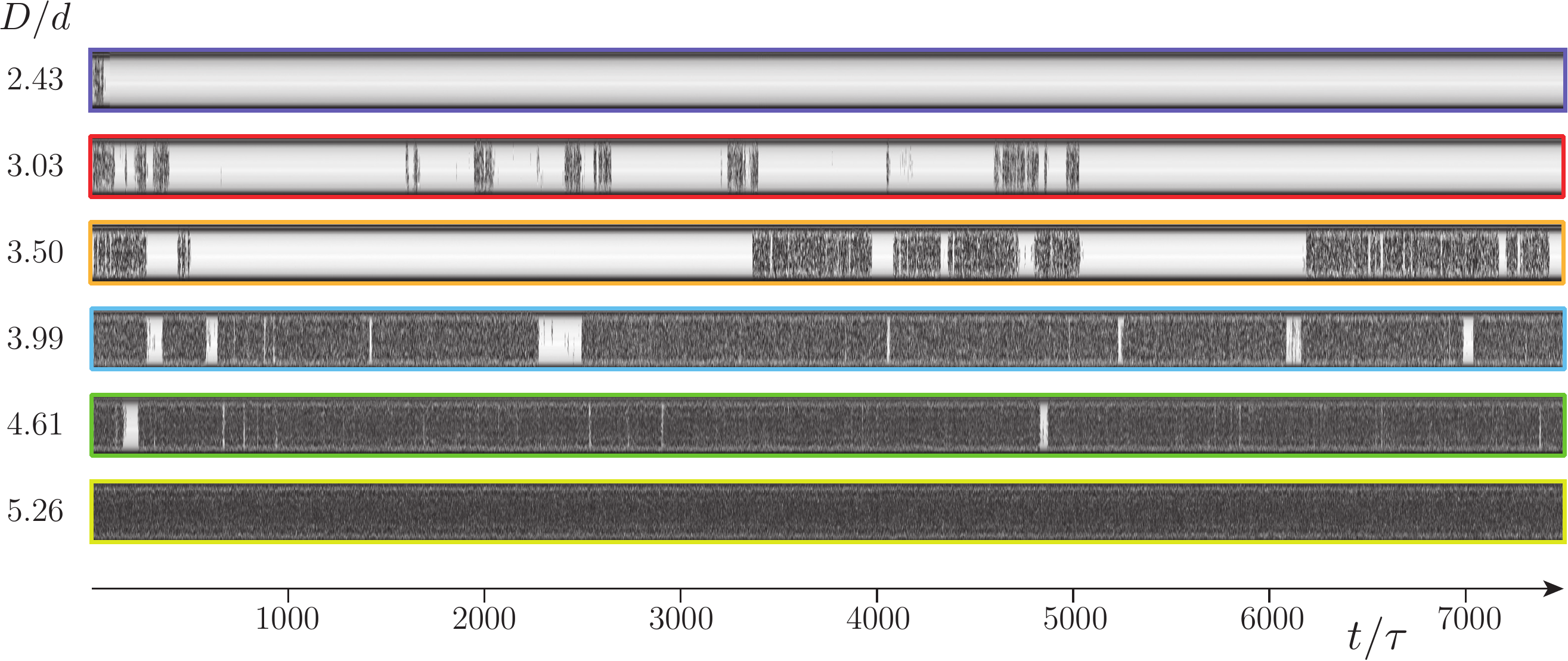}
	\caption{Spatio-temporal diagrams at the constriction neck for various $D/d$. For $D/d\leq2.43$ only few particles (which appear in dark) escape before a permanent clog is formed. When increasing $D/d$, intermittent particle flow is observed, eventually reaching particle continuous flow for $D/d\geq5.26$.}
	\label{fig:SpatioTemp}
\end{figure*}

Among the different driving forces pushing \emph{entities} through bottlenecks, the drag imposed by liquid flow is particularly interesting for small particles. Consequently, one can find several recent studies on the flow of constricted suspensions \cite{Dressaire2017, Agbangla2012, Genovese2011, Hong2017emulsions} with applications to drug delivery \cite{Teymoori2005}, lifetime reduction of filters in separation processes \cite{Leverenz2009}, pollutant removal using subsurface treatment \cite{Knowles2011}, or in technological applications relying on the suppression of any kind of blockage such as ink-jet printing \cite{Wyss2006}. The addressed question often concerns clogging prevention, nonetheless the clogging itself can also be advantageous, for example in medicine to provoke embolization of blood vessels in order to shrink a tumor \cite{Laurent2007}. Interestingly, in spite of these numerous studies, the limits of the clogging/flowing regimes of constricted particle suspensions has received little attention, with some exceptions in other similar configurations \cite{stoop2018} or using numerical simulations \cite{Hidalgo2018}. It is indeed a complex task to tackle experimentally given the requirement of high-reproducibility, high time-resolution and large data sets for good statistics. The problem is however crucial since it would clarify the role of the driving force, and in this case, the effect of the interstitial fluid in the clogging/unclogging process, which is a matter of current active debate \cite{Marin2018, Koivisto2017, Kulkarni2010}. In this work, we explore the critical sensitivity of the intermittent particle flow on the neck-to-particle size ratio $D/d$ by using high-resolution and high-speed optical video microscopy.

%===============================================================================
%	SETUP
%===============================================================================
%\section{Experimental set-up}\label{exp}

A schematic of the experimental set-up is shown in Figure \ref{fig:setup}a. The fluidic system consists of a transparent straight channel of borosilicate glass (isotropic wet etching fabricated by Micronit microfluidics) 
{\color{black} with a rectangular cross-section of $100 \,\times\, 400\,\SImum^2$ which reduces to an almost square cross-section of $100\,\times\,110\,\SImum^2$ to form the neck. The constriction is achieved by a linear narrowing of the channel with a half-angle of 60\degree. This specific design, similar of that used in Marin et \emph{al.} \cite{Marin2018}, forms a two-dimensional nozzle converging towards the neck.
}
%of uniform thickness 100 $\SImum$, which is locally constricted in the middle. The constriction is achieved by a linear narrowing of the channel with a half-angle of 60\degree~from the nominal width of 400 $\SImum$ down to 110 $\SImum$ at the neck. This specific design, similar of that used in Marin et \emph{al.} \cite{Marin2018}, forms a two-dimensional nozzle converging towards the almost squared cross-section of the neck (110 x 100 $\SImum^2$).
Particles and liquid have been carefully chosen to avoid buoyancy effects, particle aggregation and particle deposition at the micro-channel walls, as other studies have previously analyzed \cite{Dressaire2017,Cejas2017}. The suspension consists of monodisperse {\color{black} spherical} polystyrene particles of diameter $d$ which is varied from 19.0 to 41.1 $\SImum$ (see table \ref{tab:alpha}). Particles are stabilized with negatively charged sulfate groups (Microparticles GmbH) in a density-matched 26.3 wt\% aqueous solution of glycerine, with a density $\rho=1062\,$kg/m$^3$ and a viscosity $\mu=2.1\,$mPa.s. The charged sulfate groups confer them a small negative surface potential (on the order of $-50$ mV) but sufficient to prevent both their agglomeration and their adhesion to the channel walls. The suspension is driven through the constriction at constant volume flux using a syringe pump (Harvard Apparatus), at a Reynolds number $Re=\rho \langle U\rangle d /\mu\sim10^{-3}$ (corresponding to particle velocities $\langle U\rangle\approx 5\,$mm/s). The average particle velocity at the constriction neck $\langle U \rangle$ is used to define the Stokes time $\tau=d/\langle U\rangle$ (the time a particle takes to travel its own diameter), which will be used as the characteristic time scale. {\color{black} The use of a syringe pump to drive the liquid flow implies that a constant liquid volume flux is forced through the channel, even when an arch forms at the bottleneck blocking particles. In that case, only liquid flows through the bottleneck and pressure increases due to the increased hydrodynamic resistance. In some cases, the pressure growth may reach up to 7 bars; if this happens, the experiment is automatically stopped and re-initialized to avoid damaging the setup. Varying the flow rate within the range of $Re \ll 1$ that we have explored does not change the results. Note also that the} reason for using such particle size range (between 10 and 50 $\SImum$) is dual: on the one hand, we avoid colloidal particle interactions and Brownian motion. On the other hand, increasing the particle size also involves the flow control of larger volumes of fluid, larger $Re$ numbers and higher working pressures. Therefore, the range of particle size chosen allows us to work with highly monodisperse particles interacting mainly by lubrication interactions and low-pressure solid contacts, manipulated via microfluidic technology, which allows us to obtain a high degree of control and reproducibility difficult to achieve at other length scales. The suspension is prepared with a particle volume fraction of about $2\%$, then inserted in the device and driven downstream the constriction towards a filter which only allows the fluid to flow through. Particles are therefore concentrated in that position. \\

The experiment starts when the flow is reversed and particles are dragged by the fluid towards the constriction. The particle volume fraction as the particle reach the constriction is $\approx60\%$, and the suspension is imaged with a high-speed CMOS camera (PCO.dimax CS1, 12 bit, 1000fps) coupled to an inverted microscope (Nikon Instruments, Eclipse TE2000-U). Figure \ref{fig:setup}b presents successive snapshots of a typical intermittent flow experience (see also Movie 1 \cite{ref-Movie1}). In particular, the flow becomes interrupted by the spontaneous formation of arches spanning the bottleneck (top panel in Figure \ref{fig:setup}b). At this moment, the fluid keeps passing through the particles interstices perturbing the arches which may eventually collapse. If this happens, the flow of particles is resumed (middle panel in Figure \ref{fig:setup}b) and a burst develops until a new clog arrests the flow again (bottom panel). 
It is important to note that arches are stabilized by the mutual friction among particles, implying that despite electro-static stabilization particle-to-particle contact is unavoidable in our system. However, these contacts are not long-lasting as particles separate downstream the constriction upon resuming the flow.

The overall intermittent behavior, in which several flow and arrested periods of time alternate, can be better visualized in the spatio-temporal diagrams in Figure \ref{fig:SpatioTemp}. There, in order to analyze the different regimes of particle flow, we report the results obtained for particle diameters $d = 41.1, 33.0, 28.6, 25.1, 21.7\,\rm{or}\,19.0\,\SImum \,(\pm3\%$). Adopting the neck height $D=100\,\SImum$ as the characteristic length scale, these correspond to neck-to-particle ratios $D/d= 2.43, 3.03, 3.5, 3.99, 4.61,\,\rm{and}\,5.26$, respectively (see table \ref{tab:alpha}). A spatio-temporal diagram is constructed by selecting a vertical line of unit pixel width at the middle of the constriction. This line is then sampled for every frame and is stacked alongside (see also Movie 2 \cite{ref-Movie2}). The diagrams clearly reveal a qualitative difference in the flow behavior: from a blocked situation (top diagram) to uninterrupted particle flows (bottom diagram). For $D/d\leq2.43$ only few particles (which appear in dark) escape before a clog is formed. As $D/d$ is increased, the intermittent particle flow regime emerges and the flowing intervals become longer and more abundant. This regime persists until the flow becomes continuous, so the bursts intermittency becomes unmeasurable. Note that in this system, the intermittent flow is only observed for $2.43<D/d<5.26$. We must emphasize that with a similar setup, Marin \emph{et al.} \cite{Marin2018} reported continuous flow for neck-to-particle size ratio down to $D/d=4$. Although their observation was not based on a systematic study, it is consistent with the results reported here if we consider that the constriction angle in their setup was smaller (45\degree) than the one consider here (60\degree). In a silo full of dry grains, reducing the hopper angle (with respect to the vertical direction) has been shown to decrease clogging probability in such quantitative way \cite{To2001,Lopez2019}.

\begin{table}[]
\begin{tabular}{c|c|c|c|c|c|c}

D/d                   & 2.43             & 3.03 & 3.5  & 3.99 & 4.61 & 5.26            \\ \hline
d ($\SImum$)                     & 41.1             & 33.0 & 28.6 & 25.1 & 21.7 & 19.0            \\ 
\hline
$\alpha$ & {\footnotesize Permanent clog} & 1.46 & 1.48 & 1.84 & 1.94 & {\footnotesize Continuous flow}

\end{tabular}
\caption{Values of alpha for various neck-to-particle size ratios}
\label{tab:alpha}
\end{table}

\begin{figure}
	\centering
	\includegraphics[trim =30 0 0 0,width=0.45\textwidth]{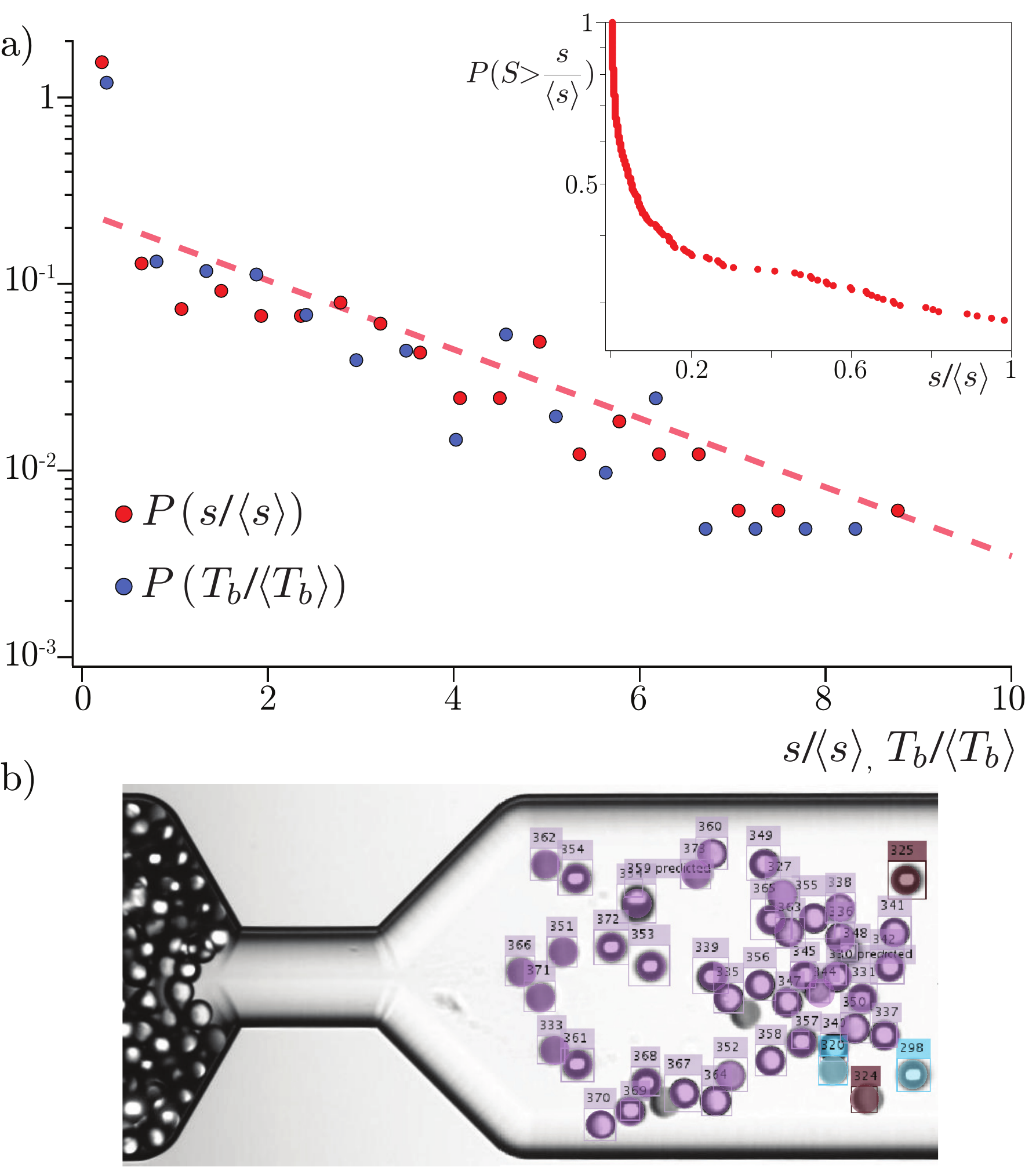}
	\caption{a) Distribution of the normalized number of escaped particles $s/\langle s\rangle$ and burst duration $T_b/\langle T_b\rangle$ for $D/d=3.03$, for which $\langle s\rangle$=270 and $\langle T_b\rangle=55\tau$. {\color{black}The dashed-line corresponds to the expected trend $\propto e^{-s/\langle s\rangle}$. Inset: Survival function $P(S>s/\langle s\rangle)$ limited to the range $s/\langle s\rangle<0.2$. The steeper slope indicates a high probability of having small particle bursts.} b) Typical particle burst showing the particle tracking employed for counting individual particles. The colors code for particles escaping during the same burst.}
	\label{fig:Ps}
\end{figure}

In what follows, we will quantify the intermittent dynamics by analyzing separately the clog formation and clog destruction processes by looking at the statistics of burst sizes and arrest times respectively.

\emph{Clog development.} In this particular system, the formation of arches has been described using a simple stochastic model \cite{Marin2018}, considering that an arch develops when a sufficient number of randomly arriving particles reach the constriction in the appropriate arrangement (see an example of a minimalistic arch formation on Movie 3 \cite{ref-Movie3}). Similarly, Thomas and Durian \cite{Thomas2015} proposed a model for dry systems in which they showed that the discharged particle mass grows as an exponential function of hole diameter. Both models are actually compatible with the idea that new \emph{microstates} \cite{Thomas2015} in the vicinity of the constriction are continuously and randomly sampled while particles arrive, until a stable arch is eventually found. One consequence of such clogging mechanism contemplated by both models is that there is no sharp clogging transition for a given critical outlet size in the sense that there is always a non-zero probability for a clog to occur.

It is therefore important to test the stochasticity of the particle bursts: Particle clogs should occur randomly with the same probability for all particles to get trapped; consequently the number of particles per bursts should follow an exponential distribution as it does in avalanches or fractures occurring stochastically \cite{Fisher1998}. It is then expected that the distribution of particle number per burst follows an exponential distribution, as it is well-known for particles in silos \cite{Zuriguel2003}, Brownian \cite{Hidalgo2018} and non-Brownian suspensions \cite{Marin2018}, pedestrians \cite{Garcimartin2016pedestrians} and animal flocks \cite{Garcimartin2015sheep}, regardless of the stability of the arch that separates each particle burst (permanent or intermittent). Experimentally, monitoring the number of particles per burst is not a trivial task given that particles overlap frequently in this three dimensional configuration. Measurements were done for one particular case of low $D/d$ ($=3.03$), for which the amount of overlaps is lower and tracking of individual particles is possible. A particle tracking algorithm was specifically developed to accurately track every escaping particle. This method is based on a Kalman filter \cite{kalman1960new} and allows precise prediction of the expected position of tracked particles over several frames where particles are partially or totally occluded by neighboring particles, making it particularly efficient for tracking relatively dense flow of particles, as illustrated on Figure \ref{fig:Ps}b (see also Movie 4 \cite{ref-Movie4}).

Note also that given the discrete nature of the system, defining a arrest time threshold to set apart successive bursts is not straightforward. This is done by looking at the distribution of times between the passage of two consecutive particles (as it will be shown below). From those distributions, and using the Clauset-Shalizi-Newman method \cite{Clauset2009}, a characteristic minimum time is obtained $T_{\rm{min}}$. Then, a flow interruption longer than $T_{\rm{min}}$ implies the end of a burst and the beginning of another.

Figure \ref{fig:Ps} shows the distribution of the burst size $s$ normalized by the mean burst size $\langle s\rangle$. The distribution is well fitted by an exponential distribution {\color{black}(dashed line $\propto e^{-s/\langle s\rangle}$)}, which can be explained assuming there is a constant probability of clogging during the whole burst \cite{Zuriguel2003}.
It is nonetheless surprising to find a substantially higher probability for the smallest particle bursts in the distribution. This means that, relatively often, small number of particles (of the order of $\sim\langle s\rangle/10$) manage to escape through the arch without destabilizing it. {\color{black} This feature is better observed when representing the survival function of the rescaled avalanche size: $P(S>\frac{s}{\langle s\rangle})$ as shown in the inset of Figure 3. Clearly, the steeper curve obtained for small values of $s/\langle s\rangle$ demonstrates that clogging is more likely to occur after a burst of such size}. We interpret this feature as a result from the system's inherent three-dimensionality, as such behavior is not reported for two-dimensional configurations. Although the choice of a different threshold $T_{\rm{min}}$ obviously affects the particular value of $\langle s\rangle$, we have checked that the shape of the distribution is unaffected.

Apart from analyzing the burst size distributions (in number of beads) we have also computed the burst duration $T_b$ which is defined as the lapse time going from the passage of the first particle of the burst to the last one. As expected, the distributions of $T_b$ display the same features than described before; i.e. an exponential tail with a notably higher probability for short burst durations. This confirms that these two parameters are equally suitable to describe clogging development as a Poisson process.

% ----------- burst time-distribution --------------------%

\begin{figure}[t!]
	\includegraphics[width=0.5\textwidth]{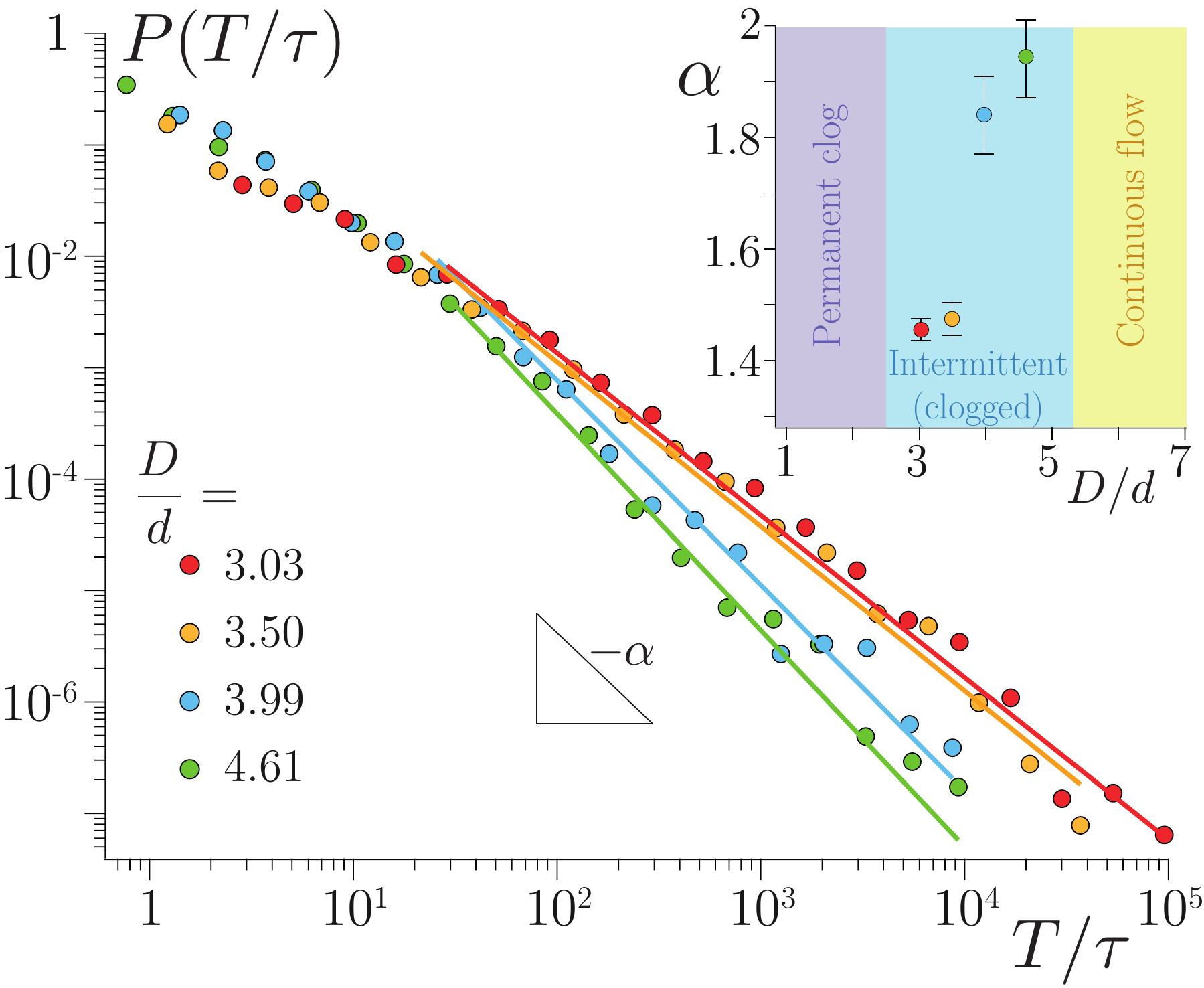}
	\caption{Distribution of the arrest time lapses normalized by the Stokes time $T/\tau$. The lines correspond to the best power law fits with their exponent $\alpha$, as given by the Clauset-Shalizi-Newman method \cite{Clauset2009}. Inset : $\alpha$ as a function of the neck-to-particle size ratio $D/d$. The exponent can only be defined when the flow is intermittent and the values obtained (always below 2) reveal that the system is in the clogged regime when $D/d<5.26$. See values of $\alpha$ in table \ref{tab:alpha}.}
	\label{fig:PTt}
\end{figure}

\emph{Clog destruction.} In order to investigate the unclogging dynamics, we proceed to analyze the probability distributions of the time lapses $T$ between the passage of consecutive particles. We have followed the same strategy implemented in previous studies on intermittent flowing systems, such as hungry sheep herds \cite{Garcimartin2015sheep}, pedestrian crowds \cite{Helbing2005, Krausz2012,Garcimartin2016pedestrians}, mice escaping a water pool \cite{Saloma2003}, or vibrated silos of dry granular material \cite{Janda2009, Lastakowski2015}. In such systems, it has been found that the distribution of arrested time lapse exhibits a power-law tail $P(T)\propto T^{-\alpha}$, thus being a signature of systems prone to clogging \cite{Zuriguel2014}. Furthermore, the value of the exponent $\alpha$ is proxy for the intermittency of the particle flux. Note that average time lapses $\langle T\rangle$ can only be defined for distributions fulfilling $\alpha>2$, while $\langle T\rangle$ diverges for $\alpha\leq2$, which we interpret as a transition to a scenario in which a permanent clog will eventually develop.  For $\alpha\leq2$, there is a non-zero probability of observing everlasting clogs, while for $\alpha>2$, the system can be temporary blocked due to the formation of clogs but no arch will persist infinitely. From our data, the exponent $\alpha$ of the power-law tail is obtained using the rigorous Clauset-Shalizi-Newman method \cite{Clauset2009}, which also yields the goodness of the fit, and gives the minimum time lapse $T_{\rm{min}}$ from which the power-law fit is valid (this is, indeed, the time lapse used to set apart consecutive bursts that was mentioned before).

Figure \ref{fig:PTt} presents the probability distribution of the arrest lapses obtained for various $D/d$ from the spatio-temporal diagrams shown in Figure \ref{fig:SpatioTemp}. The distribution $P(T/\tau)$ exhibits the characteristic power-law tail $P(T/\tau)\propto(T/\tau)^{-\alpha}$. \textcolor{black}{The time lapse probability distribution is obtained for $2.43<D/d<5.26$, when consecutive bursts are separated by measurable periods of arrest ($T\gtrsim \tau$).} Our largest measurable time-scale is limited by the maximum time that can be captured on our camera's digital memory, which sets the maximum number of frames that we can capture on each experiment. Nonetheless, note that we have been able to measure time lapses up to five orders of magnitude larger than the Stokes time. Furthermore, around 800 bursts have been analyzed for each neck-to-particle size ratio. Overall, our measured time-lapse distributions yield tails spanning at least over 3 orders of magnitude for the largest $D/d$ and up to 4 orders of magnitude for the smallest $D/d$.

As shown on the inset of Figure \ref{fig:PTt}, where $\alpha$ is plotted versus $D/d$, the exponent is remarkably sensitive to the neck-to-particle size ratio. Indeed, it goes through an abrupt increase in the range $D/d\!=\!3.50$ to $3.99$, from $\alpha\!=\!1.47 \! \pm \!0.03$ up to $1.84\pm\!0.07$. We want to stress the high sensitivity of the system on such a small increase on $D/d$, since it represents only a decrease in about 10\% on the particle size (about 3 microns decrease in particle diameter). Remarkably, for those $D/d$ for which time lapse distributions were measurable, the value of $\alpha$ \emph{exclusively} remains below 2, thus indicating a non-zero probability of observing clogs that persist indefinitely. The trend clearly suggests that the expected value for $D/d=5.26$ would lay within the area of $\alpha>2$ (non-persistent clogs). However, in this case the flow is uninterrupted, at least at the experimental time scale. This feature indicates that for $D/d$ around 5, a sharp transition occurs from a clogged state (intermittent flow with $\alpha\leq2$) to a continuous flow (zero probability of clogging or negligible in practical terms). As far as we know this is a unique property of our system. In fact, in all the other scenarios where the unclogging transition has been reported based on the power law tails of the clogging times (vibrated silos \cite{Janda2009}, simulated colloids or pedestrians \cite{Zuriguel2014}, and self propelled robots \cite{Patterson}), the passage from clogged state ($\alpha\leq2$)) to continuous flow, occurred through a region of intermittent flow with $\alpha>2$. The fact that this regime is not observed in our case may be just a consequence of the intermittent region being very narrow{\color{black}, which may be explained by the narrow size distribution of the particles used here, which reduces the number of particle configurations at the neck that can lead to arch formation}; but it can also reflect that the perturbation introduced by the fluid is not high enough to destroy the strongest arches formed in each experimental condition.

%-------------------%-------------------%-------------------%-------------------%-------------------%-------------------%-------------------

%-------------------%-------------------%-------------------%-------------------%-------------------%-------------------%-------------------

To conclude, we have shown that volume-driven flow of dense non-adhesive particle suspension going through a constriction exhibits intermittent flow behavior with a striking similarity as in dry granular matter, human crowds, or animal herds \cite{Zuriguel2014}, which we have demonstrated in this work in a quantitative manner. \textcolor{black}{At low Reynolds number and in the absence of particle aggregation or particle/wall adhesion, the geometry of the system (neck-to-particle size ratio and constriction angle) prescribes the flow regime entirely.} At low $D/d$ the flow is quickly interrupted by the formation of stable particle arches spanning the constriction, while at large $D/d$ the flow is continuous and no stable arches develop (at least in the experimental timescale). A transitioning intermittent regime is reported for $2.43<D/d<5.26$, for which we have investigated the distribution of the time lapse $T/\tau$ between the passage of consecutive particles. The distribution displays a characteristic power-law tail $\propto\!(T/\tau)^{-\alpha}$ with values of $\alpha$ below two in all cases. This unexpected result implies that, according to the definition given in \cite{Zuriguel2014}, our system exhibits a sharp transition from a clogged state to a continuous flow for $D/d\approx 5$.\\

% \begin{acknowledgments}
The authors acknowledge fruitful discussions with Pallav Kant, Angel Garcimart\'in and Raul Hidalgo. AM acknowledges support from the ERC (European Research Council) Starting Grant (grant agreement No.678573). IZ acknowledges support from FIS2017-84631-P, MINECO/AEI/FEDER, UE. 
% \end{acknowledgments}

\bibliographystyle{apsrev4-1}

\begin{thebibliography}{38}

% [23] Y. Tserkovnyak, A. Brataas, and G.
%E. W. Bauer, Enhanced Gilbert Damping in Thin Ferromagnetic Films,
%Phys. Rev. Lett. 88, 117601 (2002).

\bibitem{Zuriguel2014}
% \textsc
{I. Zuriguel, D. R. Parisi, R. C. Hidalgo \emph{et al.},}
Clogging transition of many-particle systems flowing through bottlenecks,
\emph{Sci. Rep.}, \textbf{4} 7324 (2014).


\bibitem{Garcimartin2015sheep}
% \textsc
{A. Garcimart{\'i}n, J. M. Pastor, L. M. Ferrer, J. J. Ramos, C. Mart{\'i}n-G{\'o}mez and I. Zuriguel,}
Flow and clogging of a sheep herd passing through a bottleneck,
\emph{Phys. Rev. E}, \textbf{ 91(2)} 022808 (2015).

\bibitem{Helbing2005}
% \textsc
{D. Helbing, L. Buzna, A. Johansson, and T. Werner,}
Self-organized pedestrian crowd dynamics: Experiments, simulations, and design solutions,
\emph{Transp. Sci.}, \textbf{ 39(1)} 1-24 (2005).

\bibitem{Krausz2012}
% \textsc
{B. Krausz and C. Bauckhage,}
Loveparade 2010: Automatic video analysis of a crowd disaster,
\emph{Comput Vis Image Underst}, \textbf{ 116(3)} 307-319 (2012).

\bibitem{Janda2009}
% \textsc
{A. Janda, D. Maza, A. Garcimart\'in, E. Kolb, J. Lanuza and E. Cl\'ement,},
Unjamming a granular hopper by vibration,
\emph{EPL}, \textbf{ 87(2)} 24002 (2009).

\bibitem{Zuriguel2003}
% \textsc
{I. Zuriguel, A. Garcimart\'in, D. Maza, L. A. Pugnaloni and J. M. Pastor,}
Jamming during the discharge of granular matter from a silo,
\emph{Phys. Rev. E}, \textbf{ 71(5)} 051303 (2005).

\bibitem{To2001}
% \textsc
{K. To, P. Y. Lai, and H. K. Pak,}
Jamming of granular flow in a two-dimensional hopper,
\emph{Phys. Rev. Lett.}, \textbf{ 86(1)} 71 (2001).

\bibitem{Mankoc2009}
% \textsc
{C. Mankoc, A. Garcimart\'in, I. Zuriguel, D. Maza, and L. A. Pugnaloni,}
Role of vibrations in the jamming and unjamming of grains discharging from a silo,
\emph{Phys. Rev. E}, {\textbf{ 80(1)}} 011309 (2009).

\bibitem{Sloan2011}
% \textsc
{E. D. Sloan, C. Koh, A. Sum, A. L. Ballard, J. Creek, M. Eaton and L. Talley,}
\emph{Natural gas hydrates in flow assurance},
(Gulf Professional Publishing-Elsevier, Boston, 2011).

\bibitem{Rees2012}
% \textsc
{D. G. Rees, H. Totsuji and K. Kono,}
Commensurability-dependent transport of a Wigner crystal in a nanoconstriction,
\emph{Phys. Rev. Lett.}, {\textbf{ 108(17)}} 176801 (2012).

\bibitem{Wu1993}
% \textsc
{X. L. Wu, K. J. M{\aa}l{\o}y, A. Hansen, M. Ammi, and D. Bideau,}
Why hour glasses tick,
\emph{Phys. Rev. Lett.}, \textbf{71}(9) 1363 (1993).

\bibitem{Saloma2003}
% \textsc
{C. Saloma, G. J. Perez, G. Tapang, M. Lim and C. Palmes-Saloma,}
Self-organized queuing and scale-free behavior in real escape panic,
\emph{Proc. Natl. Acad. Sci.}, \textbf{100}(21) 11947-11952 (2003).

\bibitem{Dressaire2017}
% \textsc
{E. Dressaire and A. Sauret,}
Clogging of microfluidic systems,
\emph{Soft Matter}, \textbf{ 13}(1) 37-48 (2017).

\bibitem{Agbangla2012}
% \textsc
{G. C. Agbangla, E. Climent and P. Bacchin,}
Experimental investigation of pore clogging by microparticles: Evidence for a critical flux density of particle yielding arches and deposits,
\emph{Sep. Purif. Technol.}, \textbf{101} 42-48 (2012).

\bibitem{Genovese2011}
% \textsc
{D. Genovese and J. Sprakel,}
Crystallization and intermittent dynamics in constricted microfluidic flows of dense suspensions,
\emph{Soft Matter}, {\textbf{ 7(8)}} 3889-3896 (2011).


\bibitem{Hong2017emulsions}
{X. Hong, M. Kohne, M. Morrell, H. Wang, H., and E. R. Weeks,}
Clogging of soft particles in two-dimensional hoppers,
\emph{Phys. Rev. E}, \textbf{96}, 06260 (2017)

\bibitem{Teymoori2005}
% \textsc
{M. M. Teymoori and E. Abbaspour-Sani,}
Design and simulation of a novel electrostatic peristaltic micromachined pump for drug delivery applications,
\emph{Sensor Actuat A-Phys}, {\textbf{ 117(2)}} 222-229 (2005).

\bibitem{Leverenz2009}
% \textsc
{H. L. Leverenz, G. Tchobanoglous and J. L. Darby,}
Clogging in intermittently dosed sand filters used for wastewater treatment,
\emph{Water Res}, {\textbf{ 43(3)}} 695-705 (2009).

\bibitem{Knowles2011}
% \textsc

{P. Knowles, G. Dotro, J. Nivala and J. Garc\'ia,}
Clogging in subsurface-flow treatment wetlands: occurrence and contributing factors,
\emph{Ecol. Eng.}, {\textbf{ 37(2)}} 99-112 (2011).

\bibitem{Wyss2006}
% \textsc

{H. M. Wyss, D. L. Blair, J. F. Morris, H. A. Stone and D. A. Weitz,}
Mechanism for clogging of microchannels,
\emph{Phys. Rev. E}, {\textbf{ 74(6)}} 061402 (2006).

\bibitem{Laurent2007}
% \textsc
{A. Laurent,}
Microspheres and nonspherical particles for embolization,
\emph{Tech Vasc Interv Radiol.}, {\textbf{ 10(4)}} 248-256 (2007).

\bibitem{stoop2018}
{R. L. Stoop and P. Tierno,}
Clogging and jamming of colloidal monolayers driven across disordered landscapes,
\emph{Commun. Phys.}, {\textbf{ 1(1)}} 68 (2018).

\bibitem{Hidalgo2018}
{R. C. Hidalgo, A. Go{\~n}i-Arana, A. Hern\'andez-Puerta and I. Pagonabarraga,}
Flow of colloidal suspensions through small orifices,
\emph{Phys. Rev. E}, {\textbf{ 97(1)}} 012611 (2018).

\bibitem{Kulkarni2010}
% \textsc
{S. D. Kulkarni, B. Metzger and J. F. Morris,}
Particle-pressure-induced self-filtration in concentrated suspensions,
\emph{Phys. Rev. E}, {\textbf{ 82(1)}} 010402 (2010).

\bibitem{Koivisto2017}
% \textsc

{J. Koivisto and D. J. Durian,}
Effect of interstitial fluid on the fraction of flow microstates that precede clogging in granular hoppers,
\emph{Phys. Rev. E}, {\textbf{ 95(3)}} 032904 (2017).

\bibitem{Marin2018}
% \textsc
{A. Marin, H. Lhuissier, M. Rossi and C. J. K{\"a}hler,}
Clogging in constricted suspension flows,
\emph{Phys. Rev. E}, {\textbf{ 97(2)}} 021102 (2018).

\bibitem{Cejas2017}
{C. M. Cejas, F. Monti, M. Truchet, J.-P. Burnouf and P. Tabeling,} Particle Deposition Kinetics of Colloidal Suspensions in Microchannels at High Ionic Strength. \emph{Langmuir} \textbf{33}, 6471-6480 (2017).

\bibitem{ref-Movie1}
{See Supplemental Material at [URL will be inserted by
publisher] for a movie of a dense particulate suspension moving through a constriction for $D/d=3.03$, flowing intermittently in a succession of erratic bursts separated by short period of arrest.}

\bibitem{ref-Movie2}
{See Supplemental Material at [URL will be inserted by
publisher] for a movie highlighting how spatio-temporal plots are constructed by selecting a (white) line, sampling it for every frame and stacking it alongside.}

\bibitem{Lopez2019}
{D. L\'opez-Rodr\'{\i}guez, D. Gella, K. To, D. Maza, A. Garcimart\'{\i}n and I. Zuriguel},
Effect of hopper angle on granular clogging,  \emph{Phys. Rev. E}, {\textbf{ 99(3)}} 032901 (2019).

\bibitem{ref-Movie3}
{See Supplemental Material at [URL will be inserted by
publisher] for a movie showing an example of minimalistic arch formation where 4 particles reach the constriction neck simultaneously and form an arch.}

\bibitem{Thomas2015}
% \textsc
{C. C. Thomas and D. J. Durian,} Fraction of Clogging Configurations Sampled by Granular Hopper Flow. \emph{Phys. Rev. Lett.} \textbf{114}, 178001-5 (2015).

\bibitem{Fisher1998}
% \textsc
{D. S. Fisher,} Collective transport in random media: from superconductors to earthquakes. \emph{Phys. Rep.}, \textbf{301}, 113-150 (1998).

\bibitem{Garcimartin2016pedestrians}
% \textsc
{A. Garcimart\'in, D. R. Parisi, J. M. Pastor, C. Mart\'in-G\'omez, C. and I. Zuriguel}, Flow of pedestrians through narrow doors with different competitiveness. \emph{J. Stat. Mech.: Theor. Exp.} 043402 \textbf{(2016)}.

\bibitem{kalman1960new}
% \textsc
{R. E. Kalman}, A new approach to linear filtering and prediction problems. \emph{J. Basic Eng.} pp. 35-45 (1960).

\bibitem{ref-Movie4}
{See Supplemental Material at [URL will be inserted by
publisher] for a movie showing the particle tracking employed for counting individual particles over several successive bursts. The colors code for particles escaping during the same burst.}

\bibitem{Clauset2009}
% \textsc
{A. Clauset, C. R. Shalizi and M. E. Newman,}
Power-law distributions in empirical data,
\emph{SIAM rev}, {\textbf{ 51(4)}} 661-703 (1995).

\bibitem{Lastakowski2015}
% \textsc

{H. Lastakowski, J. C. G\'eminard and V. Vidal,}
Granular friction: Triggering large events with small vibrations,
\emph{Sci. rep.}, {\textbf{ 5}} 13455 (2015).

\bibitem{Patterson}
% \textsc

{G. A. Patterson, P. I. Fierens, F. Sangiuliano Jimka, P. G. K\"onig, A. Garcimart\'{\i}n, I. Zuriguel, L. A. Pugnaloni and D. R. Parisi,}
Clogging Transition of Vibration-Driven Vehicles Passing through Constrictions,
\emph{Phys. Rev. Lett.}, {\textbf{ 119}} 119 (2017).


%\bibitem{Haw2004}
%% \textsc
%
%{Haw, M. D.}
%Jamming, two-fluid behavior, and `self-filtration' in concentrated particulate suspensions,
%\emph{Physical Review Letters}, {\textbf 92(18)} 185506 (2004).
%
%\bibitem{Pansu1984}
%% \textsc
%
%{Pansu, B., Pieranski, P., \& Pieranski, P.}
%Structures of thin layers of hard spheres: high pressure limit,
%\emph{Journal de Physique}, {\textbf 45(2)} 331-339 (1984).



%\bibitem{Bacchin2011}
%\textsc{Bacchin, P., Marty, A., Duru, P., Meireles, M., \& Aimar, P.}
%Colloidal surface interactions and membrane fouling: Investigations at pore scale,
%\emph{Advances in colloid and interface science}, {\textbf 164(1-2)} 2-11 (2011).

\end{thebibliography}

\end{document}